\documentclass[journal]{IEEEtran}
\IEEEoverridecommandlockouts
\usepackage{cite}
\usepackage{caption,subcaption}
\usepackage{amsmath,amssymb,amsfonts}
\usepackage{algorithm}
\usepackage{algorithmic}
\usepackage{graphicx}
\usepackage{textcomp}
\usepackage{xcolor}
\usepackage{diagbox}
\usepackage{array} 
\usepackage{multirow} 
\usepackage{makecell}
\usepackage{booktabs}
\newcommand{\g}{$G(V,E)$}
\newcommand{\link}{$L(u,v)$}
\newcommand{\tr}{$TR^{t,b}_{L(u,v)}$}
\newcommand{\lsc}{$L_{sc}(u,v)$}
\newcommand{\cone}{$C^{t,1}_L(u,v)$}
\newcommand{\czero}{$C^{t,0}_L(u,v)$}
\newcommand{\cb}{$C^{t,b}_{L(u,v)}$}
\newcommand{\re}{$R^{t,b}_{L(u,v)}$}
\newcommand{\w}{$W_{L(u,v)}$}
\newcommand{\wsc}{$W_{L_{sc}(u,v)}$}
\newcommand{\q}{$Q_b$}
\newcommand{\alp}{$\alpha$}
\newcommand{\demand}{$f_{s,d}$}
\newcommand{\pdenmand}{$P_{f_{s,d}}$}
\newcommand{\ttl}{$TL^{t,b}_{L(u,v)}$}
\newcommand{\tal}{$AL^{t,b}_{L_{sc}(u,v)}$}
\newcommand{\frh}{$FR^{f_{s,d}}_{P_H}$}
\newcommand{\frl}{$FR^{f_{s,d}}_{P_L}$}

\newcommand{\wnew}{$W^{'}$}

\newcommand{\n}{$n$}
\newcommand{\weightfunction}{$$W_{L(u,v)} := Q_b \times e^{ \frac{\alpha(C^{t,b}_{L(u,v)}-R^{t,b}_{L(u,v)})}{C^{t,b}_{L(u,v)}}},\ b \in \left\{\\0,1\}\ \right. \eqno{(1)}$$}
\newcommand{\weightfunctiontwo}{$$W^{'}_{L(u,v)} := e^{|TL^{t,b}_{L(u,v)}|}\eqno{(2)}$$}
\newcommand{\linkcapacityutilization}{$\frac{C^{t,b}_{L(u,v)}-R^{t,b}_{L(u,v)}}{C^{t,b}_{L(u,v)}}$}
\newcommand{\method} {SD-FFR}
\newcommand{\fullmethod} {Software Defined Fast Failure Recovery}
\begin{document}
\title{ \method: \fullmethod\ Mechanism in the Automatic Warehouse
}

\author{
\IEEEauthorblockN{
Ting-Cian Bai and   
Chin-Ya Huang   
}         

\thanks{
T.-C. Bai and C.-Y. Huang are with the Department of Electronic and Computer Engineering, National Taiwan University of Science and Technology, Taiwan.
(e-mail: \{M10802214, chinya\}@gapps.ntust.edu.tw)}
}

\maketitle

\begin{abstract}
Due to the rapid development of IoT technology, automatic guided vehicles (AGVs) interact with an industrial control system (ICS) through the wireless network to support the freight distribution in the automated warehouse.  
However, the message exchange among AGVs and the ICS would experience large packet loss or long transmission delay, 
in the presence of wireless network link failure and link congestion. 
Therefore, the performance of warehouse automation would be degraded.
In this paper, we propose the \fullmethod \ (\method) mechanism, aiming to improve network reliability and avoid link congestion caused by load unbalance.
With \method, when link failure occurs, the SDN controller actively detects the link failure within milliseconds, and then reroutes the affected traffic flows accordingly. 
The proposed \method \ mechanism also takes load balance into consideration in selecting routing paths when a new traffic flow joins or link failure occurs in the network.
The evaluation results show that network performance can be time efficiently improved.
\end{abstract}


\section{Introduction}
To support large number of mobile devices working in the automatic warehouse, WiFi is considered to provide network services \cite{7968828} because of its low cost comparing to 4G/5G.
In this way, the automatic guided vehicles (AGVs) in the automatic warehouse can easily communicate with the industrial control system (ICS) to exchange maintenance messages, even when they are on the move.
For example, a wireless network architecture, Industrial Automation-Factory Automation (WIA-FA), is designed and implemented in factory automation \cite{7968828}. 
Furthermore, some researches also plan differently to react to different AGV usage scenarios in the industrial networks such as  \cite{7991971}.
A flexible and reliable network is designed in the warehouse shuttle system by IEEE 802.15.4 to support dense goods placement in \cite{7991971}.
Authors in \cite{inproceedings} combine the advantages of two network architecture, wireless mesh network and the enterprise network, to 
improve the performance of AGVs in transmitting unicast and multicast traffic. Specifically, the mesh capability is applied when applicable.
Additionally, existing wireless communication technologies are investigated in satisfying the various latency and reliability requirements of data transmission
to assist the use cases of AGV and the unmanned vehicle in industrial 4.0 \cite{8539695}. 
Under this condition, the messages are exchanged through wireless links in the network. However, a link may be congested if too many messages are forwarded  through it. On the other hand, a link may fail to transmit message when fault occurs. 
When a link encounters congestion or failure, the message exchange may experience of loss or delay which in turn would introduce incorrect decision making and significantly degrade the warehouse operation.
Specifically, to ensure the AGVs actually perform the required tasks in the automatic warehouse, the interruption or long delay in message exchange is not allowed \cite{8884675}.

To reliably and flexibly route traffic flows between the AGVs and the ICS, in this paper, we consider dual-band WiFi access points (APs) to form
a network consisting of wired and wireless connectivity to route packets. Specifically, APs create a wireless mesh network (WMN) \cite{7566486}
by wirelessly connecting with each other.
Further, the software defined network (SDN) \cite{8249869} is applied in the network to dynamically route traffic flows according to the network status.
SDN is a network architecture decoupling the network control plane and the data plane of a switch, and then
a centralized SDN controller is introduced to control the packet forwarding in the whole network. 
In this way, the network management can be done globally. 
Considering the integration of SDN and WMN, authors in \cite{7536958} 
adopt the OpenWrt framework on the development board aiming to build the real network environment of SDN integration with WMN architecture and determines the feasibility of SDN and WMN operation in the real network environment.  
In \cite{7778610,7577087}, authors 
integrate WMN and SDN so that modularized and flexible routing strategies can be provided through SDN. 
In this way, the network performance can be enhanced because the strategies are designed to take advantages and eliminate the limitations of WMN by SDN.
Wireless Mesh Software Defined Networks (wmSDN) combines the SDN and WMN aims to balance traffic loads on each wireless link which in turn enhances the user performance \cite{6673345}.

In SDN, two types of failure recovery strategies, active and passive strategy, are considered.
On the other hand, the passive strategy reactively adjusts routing decisions when link failure occurs, while the active strategy pre-deploys backup resources in the network.
For example, the shared ring method is proposed in \cite{8406152} to reuse backup resources, aiming to cost effectively pre-deploying backup resources in TCAM.
Although both recovery delay and the TCAM usage could be improved, the congestion may occurs in the post-recovery network. 
A particular method of the active recovery strategy is also developed in \cite{8809178}
by using RESTful API to per-forward the redundant rules periodically. Thus, the network services can be actively recovered, when the fault occurs. 
However, the backup rules are pushed periodically, therefore the failure may not be recovered immediately if the latest rules are not pushed.
On the other hand, in \cite{8460087},  a reactive failure recovery module in multi-radio multi-channel software-defined WMNs is design and implemented to improve failure recovery time. 
However, authors do not consider the computation delay and communication delay in finding rerouting paths. 
Authors in \cite{8690787}  utilize dual-band connectivity of an AP to enhance the failure recovery efficiency. However, the performance may be limited, because the AP takes care of the link failure detection and recovery in addition to packet forwarding. Further, load balance is not considered in the post-recovery network. 

To provide a flexible failure recovery scheme, this paper adopts the passive strategy for the failure recovery. 
However, the computational complexity in calculating a rerouting path and the overhead in message exchange between the SDN controller and switches would limit the recovery performance.
In \cite{7779007}, authors develop
the local fast reroute algorithm (LFR) scheme for traffic aggregation when it occurs link failure. If  link failure is detected in the LFR, all flow aggregate to a single flow and deploy rerouting paths by the SDN controller in order to minimize the utilization of flow entries. 
But, authors do not consider the load balance which might introduce congestion in post-recovery condition.
A 
Priority-based flow control (PFC) is proposed in \cite{8528499} by designing a variety of flow priorities to reduce the flow processing delay between the SDN controller and switches. 
However, LFR is not mainly designed to support reliability. 

In this paper, we propose a \fullmethod\ (\method) mechanism to adaptively route traffic flows in the network.  
With \method, the SDN controller utilizes the available 
link bandwidth to make routing decisions when network condition changes.
Four modules, ``network resource monitoring module'',``link weight management module'',``routing decision module'' and ``failure recovery module'' are introduced and integrated carefully in \method.
When a new flow joins the network, the SDN controller updates the link weights according to the available link bandwidth, before making routing decisions aiming to avoid link congestion.
The SDN controller also updates the link weights after link failure recovered. 
Moreover, to time efficiently recover link failure, the SDN controller maintains two type of flow entries, high-priority flow entries and  low-priority flow entries implying a backup routing path of each traffic flow is selected in the SDN controller beforehand.
When a link fails, the SDN controller immediately provides rerouting information to the network, based on the link failure information and the high-priority flow entries.
Hence, the proposed \method\ not only time efficiently recovers link failure but also better balances traffic loads in the network.

The remainder of this paper is organized as follows. Section \uppercase\expandafter{\romannumeral2} presents system architecture. Section \uppercase\expandafter{\romannumeral3} describes the method proposed in this paper. 
 Section  \uppercase\expandafter{\romannumeral4} and \uppercase\expandafter{\romannumeral5} shows 
 the implementation in commercial APs and the experimental results, respectively. 
We conclude this paper 
in Section \uppercase\expandafter{\romannumeral6}.

\section{System Architecture}
\subsection{System Description}
\begin{figure}[htbp]
\renewcommand{\figurename}{Fig.}
\centerline{\includegraphics[width=0.95\columnwidth]{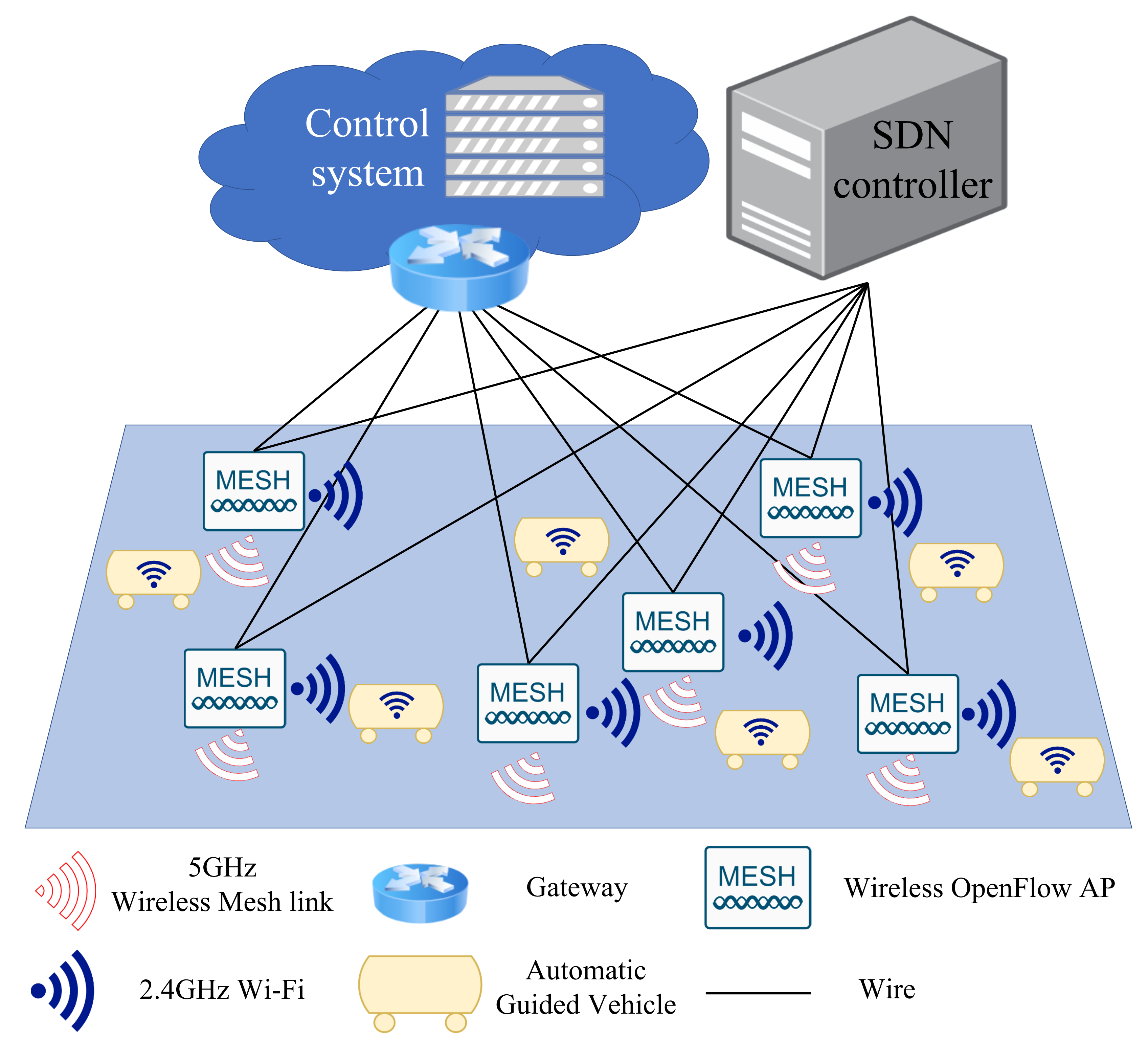}}
\caption{Illustration of SDN assisted wireless network in the automatic warehouse.} 
\label{fig:topo1} 
\end{figure}

Fig.~\ref{fig:topo1} illustrates the SDN assisted wireless network in the automatic warehouse.
Several SDN enabled dual-band wireless APs are deployed in the network to provide the wireless server to the AGVs via the 2.4 GHz frequency band.
All APs connect to the gateway through the Ethernet, and two APs in the network connects with each other via the 5 GHz wireless link to form a wireless mesh network. 
In this way, each AGV can communicate with the ICS via the gateway along the 2.4 GHz wireless link and the Ethernet.
Each AGV can also communicate with others for collaboration along the 2.4 GHz wireless link and the 5 GHz wireless mesh connectivity.
Moreover, following the OpenFlow protocol \cite{pub.1034503230}, each AP connects to the SDN controller through the Ethernet, coordinates with the SDN controller to support packet transmission in the network. 


However, since AGVs may not uniformly distributed in the network, the link congestion may occur when the traffic load on a link is too heavy. In this way, the end-to-end data transmission between the AGVs and the control system may experience long delay or packet loss.
On the other hand, the connection of a link may fail and then packets forwarding through the link cannot be delivered to the corresponding destinations successfully.
More specifically, we focus on overcoming the following two problems in degrading the network performance in this paper:
\subsection*{Problem 1: Link Failure}
When the Ethernet link failure occurs between the ICS and an AP, 
the ICS and the corresponding AGVs cannot exchange messages with each other which cause incorrect operations of the AGVs.
The performance in collaboration among AGVs may degrade when the link failure occurs at a 5 GHz wireless link connecting two APs which in turn causes the loss of delivering the synchronization messages among the collaborated AGVs. 

\subsection*{Problem 2: Link Congestion}
Typically, routing paths are chosen according to the minimum number of hops.
However, each AGV  moves dynamically in the warehouse so that teach AP may serve significantly different amount of AGVs for packet transmission.
Under this circumstance, the traffic congestion may occur at some links when the load on that link is heavy.
Moreover, in the presence of link failure, some flows are rerouted in order to sustain the message exchange between the ICS and some AGVs or exchange among the collaborated AGVs. As a consequence, some APs would degreade the performance in forwarding both the current traffic flows and the rerouted flows simultaneously due to the limited link bandwidth. 



In this way, the goal of this paper is to utilize SDN to design a scheme to reliably and efficiently route packets in the network. To focus on addressing the above mentioned problems, we in further model the network as shown in Fig.~\ref{fig:topo0} and introduce several parameters to ease of this description of the rest of the paper.

\subsection{Network Model}
\begin{figure}[h]
\renewcommand{\figurename}{Fig.}
\centerline{\includegraphics[width=0.95\columnwidth]{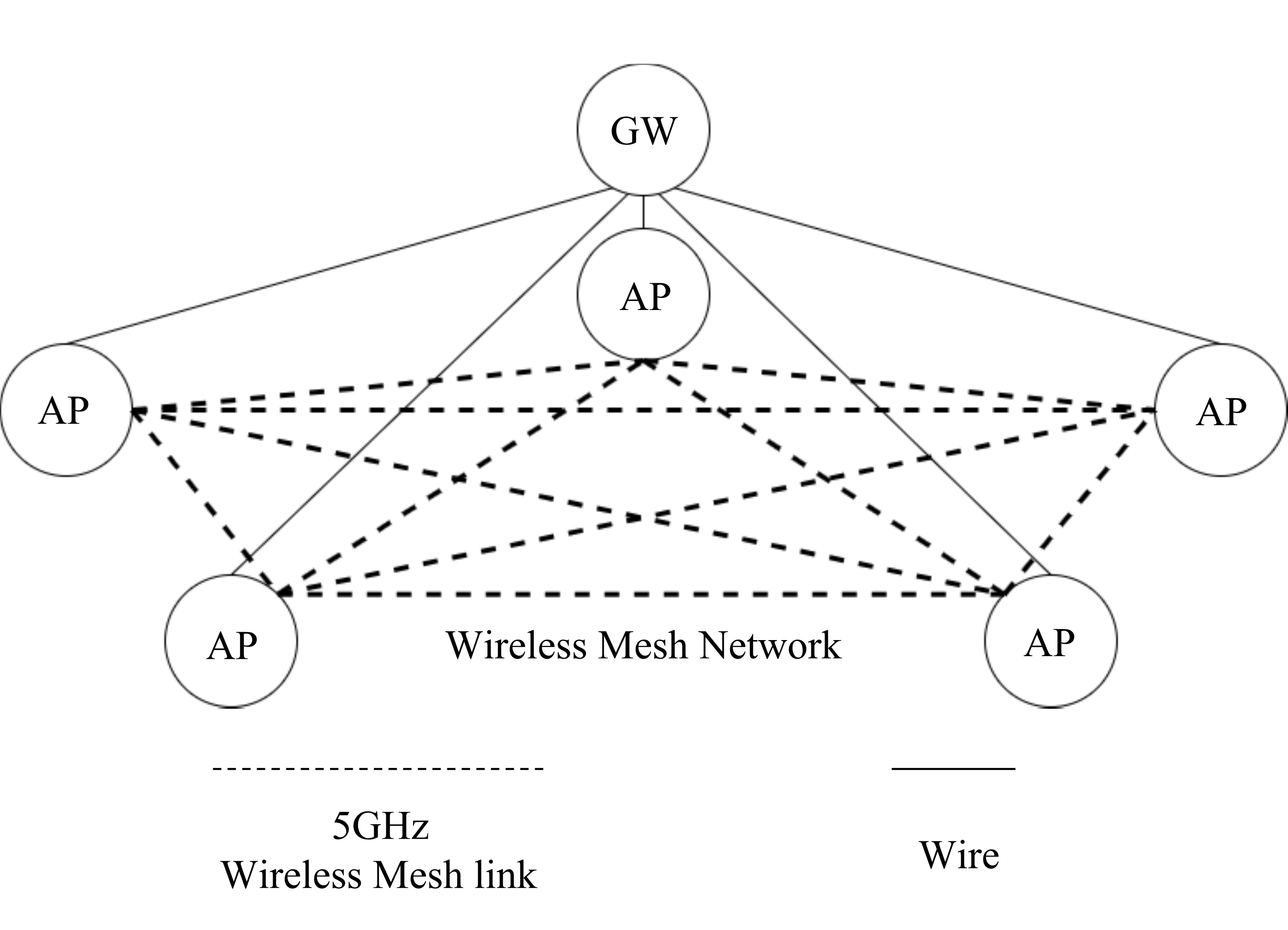}}
\caption{Network model.} 
\label{fig:topo0}
\end{figure}

We consider a network model presented in Fig.~\ref{fig:topo0} which is 
treated as an undirected graph \g. In the graph \g,  
we denote $n$ as the node,  any APs or the gateway (GW),   $V$ as 
the set of nodes, and $E$ as the set of links.
Two types of links, Ethernet and 5 GHz wireless connection are available in the topology. The GW and each AP connects with each other through the Ethernet link. Each AP connects with each other through the 5 GHz wireless link, and the 5 GHz connectivity forms a wireless mesh network. 
Furthermore, we denote the link to be \link \ $\in E$, and $u$, $v$ $\in$ $V$ denote any of the adjacent nodes in the topology. 
The link bandwidth at time $t$ is denoted to be $C_{L(u,v)}^{t,b}$, where $b$ is set to be 1 and 0, if the link is created by wireless and wired medium, respectively.  
Hence, \cone \ is the capacity of 5 GHz wireless link between two APs and \czero \ is the wired link between an AP and the GW. 
Moreover, the traffic flow sending from a source node $s$ to a destination node $d$ via this network is denoted as \demand.
Since more than one traffic flows may share \link\ to the corresponding destinations, the total amount of the traffic loads
on \link\ is denoted as \tr, where $b$ is set to be 1 and 0 for wireless and wired link, respectively. We in further denote the residual bandwidth on \link\ to be \re, and $R^{t,b}_{L(u,v)}=C^{t,b}_{L(u,v)}-TR^{t,b}_{L(u,v)}$, where $b$ is set to be 1 and 0, if the link is created by wireless and wired connection, respectively.  Moreover, the total number of traffic flows on \link\ is denoted as $|TL_{L(u,v)}^{t,b}|, b \in \{0,1\}$. Specifically, \ttl\ is denoted to be the set flows routed through each link on the network.

\subsection{Problem Formulation}
To deal with the dynamical change of traffic loads as well as the link failure in the network, we design a cost function, 
We denote $Load(t)$ to be the maximum load of a link in the network and $RD(t)$ to be failure recovery delay.
The recovery delay depends on the summation of  the processing time of the switch $H_s(t)$, the processing time of the controller $H_c(t)$, and the communication delay between the controller and switch $RTT_{s,c}(t)$. 
Moreover, when link failure occurs at \link, we set link capacity as zero, $C^{t,b}_{L_{sc}(u,v)}=0$.
Then, the total number of traffic flows and the traffic loads on \link\ are also zero, since no flows could forwarded on \link.
We have $|TL^{t,b}_{L_{sc}(u,v)}|=0$ and $TR^{t,b}_{L_{sc}(u,v)}=0$.
Consequently, the problem can be formulated as
$$\min (Load(t)+RD(t))$$
$s.t.$
$$Load(t) = \frac{\max {|TL^{t,b}_L(u,v)|}}{avg(\sum_{L(u,v)\in E}|TL_{L(u,v)}^{t,b}|)}, L(u,v)\in E$$
$$RD(t) = H_s(t)+H_c(t)+RTT_{s,c}(t)$$
$$C^{t,b}_{L(u,v)}\geq R^{t,b}_{L(u,v)}\geq TR^{t,b}_{L(u,v)}, \forall L(u,v)\in E,\ b \in \left\{\\0,1\}\ \right.$$

The objective function $\min (Load(t)+RD(t))$ 
aims at minimizing the maximal traffic loads on a link and the failure recovery delay. 
Thus, traffic flows would be time efficiently rerouted without introducing additional link congestion, 
when a traffic flow joins the network or a link failure occurs at time $t$.

Additionally, TABLE~\ref{parameter} summarizes the definition of all the network parameters using in this paper.

\begin{table}[htbp]
  \centering
  \caption{Parameter}
  \label{parameter}
  \begin{tabular}{p{2cm}||p{6cm}}
      \toprule 
      \textbf{Notation} & \textbf{Description}\\
      \midrule
      \toprule
      \g & The graph of the network topology.\\
      \link & The link between node $u$ and node $v$.  \\
      \tr & Traffic loads pass through \link \  at $t$ time instant.  \\
      \n & denote the OpenFlow device.  \\
      \lsc & The link with state change between node $u$ and $v$. \\
      $b$ & $b$ is a Binary value, $b=1$ denote the wireless mesh link  and if $b=0$ denote the Ethernet link.\\
      \cb & $C_{1}$ denote the Wireless Mesh link capacity and $C_{0}$ denote the Ether link capacity at $t$ time instant.  \\
      \re & Residual link capacity of \link \ at $t$ time instant.  \\
      \w & The weight of \link. \\
      \wsc & The weight of state change \link.  \\
      \q & $Q_{1}$ denote the adjustable value for Wireless Mesh link, and $Q_{0}$ denote the adjustable value for Ether link. \\
      \demand & Traffic demand from node $s$ to node $d$.  \\
      \pdenmand & Path of traffic demand \demand.  \\
      \ttl & The set of \demand \ pass through the link \link \ at $t$ time instant. \\
      \tal & The set of \demand \ affected by fault pass through the link \lsc \ at $t$ time instant.\\   
      \frh & High-priority forwarding rule.\\
      \frl & Low-priority forwarding rule.\\
      \bottomrule
  \end{tabular}
\end{table}

\section{\fullmethod (\method) Mechanism}

In this paper, we utilize SDN and then propose the \fullmethod \ (\method) mechanism to time efficiently react to the change of network status by conditionally reroute traffic flows. 
With \method, the performance degradation resulted from link failure and link congestion in the SDN assisted wireless network can be solved. In this way, the reliability and load balance can be properly sustained in the network. 

\begin{figure}[htbp]
\renewcommand{\figurename}{Fig.}
\centerline{\includegraphics[width=0.95\columnwidth]{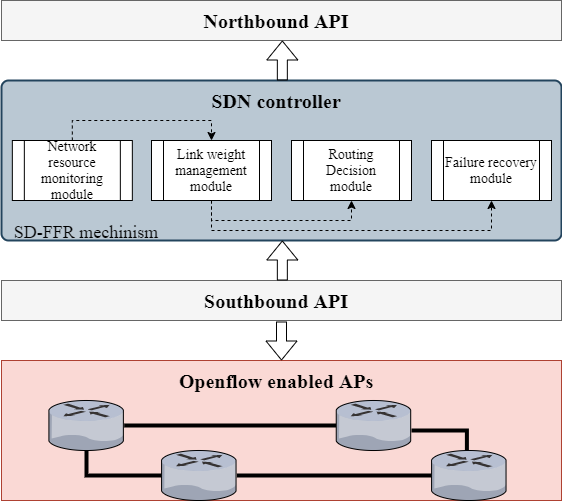}}
\caption{The architecture of SD-FFR mechanism.} 
\label{system}
\end{figure}

Fig.~\ref{system} shows the architecture of integrating the proposed \method\ mechanism into existing SDN architecture. 
Specifically, the proposed \method\ mechanism is implemented in the SDN controller. Following the OpenFlow protocol, the network status can be delivered through the Southbound API from the APs to the controller. The routing decisions are also delivered from the controller to the APs via the Southbound API. In this way, packets can be adaptively routed in the network according to the network conditions.
As illustrated in Fig.~\ref{system}, four modules consist of the \method\ mechanism, and they are ``network resource monitoring module'', ``link weight management module'', ``routing decision module'' and ``failure recovery module''.

\subsection{Network Resource Monitoring Module}
To effectively utilize the available link bandwidth in the network, the ``network resource monitoring module'' collects the information of the  available link bandwidth in data plane, 
when a traffic flow \demand \ joins the network.
Then, the collected information is passed to the ``link weight management module'' for further processing. 

Alg. 1 describes the operation of the ``network resource monitoring module''.
The main purpose of Alg. 1 is to update the current residual link bandwidth, denoting as \re, 
by collecting the traffic loads on each link, denoting as \tr, of the network, when a new flow joins the network. In this way, the updated \re\ can be used in ``routing decision module'' for routing decision making. In Alg. 1, when a new flow joins the network.

\begin{algorithm}[htbp]
  \caption*{\textbf{Alg. 1:}  Network Resource Monitoring Module}
  \begin{algorithmic}[1]
      \STATE SDN controller monitors \tr, $\forall$ $L(u,v)\in E$. 
      \STATE \textbf{If} \demand \ joins the network \textbf{then}
      \STATE \ \ \ \textbf{Forall} $\forall$ $L(u,v)\in E$ \textbf{do}
      \STATE \ \ \ \ \ \ $R^{t,b}_{L(u,v)} := C^{t,b}_{L(u,v)}-TR^{t,b}_{L(u,v)}$\\
      \ \ \ \textbf{end}\\
      \textbf{end}\\
      \STATE \textbf{Back to line 1}
    \end{algorithmic}
\end{algorithm}

\subsection{Link Weight Management Module}
To balance traffic loads in the network, the link weight management module
maintains the weight of each link.
Specifically, we introduce two parameters, $W_{L(u,v)}$ and $W^{'}_{L(u,v)}$, which are denoted to be the weights in the normal network state and the post-recovery network state, respectively.
The normal network state implies no link failure occurs while the post-recovery network state implies the flows forwarded through failed links have been recovered.
Furthermore, $W_{L(u,v)}$  is updated according to the link capacity usage, and $W^{'}_{L(u,v)}$ is adjusted based on traffic demands on each link. 

\begin{itemize} 
\item 
Link weight in normal network state: \\
\weightfunction
The link weight in the normal network state is calculated according to (1). The link weight is applied for making routing decisions to balance traffic flows and prevent traffic congestion in the network.
Hence, the available bandwidth on each link, \linkcapacityutilization.
Further, two parameters \q\ and \alp\ are set to increase the probability in selecting the wired link for data transmission aiming to utilize the large link bandwidth of the Ethernet which is usually larger than the 5 GHz wireless link.
The \q \ is in further set as $0<2Q_{0}<1$ and  $Q_{0}+Q_{1}=1$.

\item 
Link weight in post-recovery network state:\\
\weightfunctiontwo
The link weight in the post-recovery is calculated according to (2).
The link weight is applied for making routing decisions to in further balance traffic loads, after the link failure recovery.
To time efficiently adjust the link weight to assist the fast failover and the load balance, only total number of traffic demands on each link, $|TL_{L(u,v)}^{t,b}|$, is considered for the link weight calculation, 
instead of considering the traffic loads on each link as (1). 
In this way, not only the time spending in calculating the link weight can be reduced, the time spending in exchanging control messages between APs and the controller can be avoided.
\end{itemize}

The detail operation of the ``link weight management module'' is described in Alg. 2. When the network starts, the link weight, $W_{L(u,v)}$, is initialized as $Q_b$. As long as a new flow joins the network, $W_{L(u,v)}$ is updated according to the link bandwidth utilization. On the other hand, the link wight after link failure recovery, $W^{'}_{L(u,v)}$, is calculated based on the number of flows on $L(u,v)$.

\begin{algorithm}[htbp]
  \caption*{\textbf{Alg. 2:} Link Weight Management Module}
  \begin{algorithmic}[1]
      \STATE \textbf{If} Network initialization \textbf{then}
      \STATE \ \ \ $W_{L(u,v)}:=Q_b$
      \STATE \textbf{Else if} Normal network state  \textbf{then}
      \STATE \ \ \ $W_{L(u,v)} := Q_b \times exp( \frac{\alpha(C^{t,b}_{L(u,v)}-R^{t,b}_{L(u,v)})}{C^{t,b}_{L(u,v)}})$
      \STATE \textbf{Else if} Post-recovery network state  \textbf{then}
      \STATE \ \ \ $W^{'}_{L(u,v)} := exp(|TL^{t,b}_{L(u,v)}|)$\\
      \textbf{end}
    \end{algorithmic}
\end{algorithm}

\subsection{Routing Decision Module}
When the traffic demands join the network, the link weights are firstly updated, and then the ``routing decision module'' adjusts the routing path based on the Dijkstra algorithm \cite{10.1145/321765.321768} as well as installs the forwarding rule to the corresponding APs to create the routing path for the new joining flow \demand.

Alg. 3 describes the process of ``routing decision module''. 
When \demand\ joins the network, considering the updated link weights through the Alg. 2, 
the routing path \pdenmand\ in forwarding data of \demand\ is calculated through the Dijkstra algorithm. 
Then, the SDN controller adds the low-priority forwarding rule \frl \ to the corresponding nodes on path \pdenmand. 
Moreover, this algorithm maintains the set flows routed through each link on the network which is denoted as \ttl.
Specifically, \ttl\ is managed to assist the operation of ``failure recovery module''.
\begin{algorithm}[htbp]
  \caption*{\textbf{Alg. 3:} Routing Decision Module}
  \begin{algorithmic}[1]
      \STATE \textbf{Input:} Traffic demand \demand
      \STATE \textbf{Output:} Routing decision
      \STATE \textbf{If} \demand \ joins the network \textbf{then}

      \STATE \ \ \ \textbf{Forall} $L(u,v)\in E$ \textbf{do}
      \STATE \ \ \ \ \ \ \textbf{Run Alg. 2}\\
      \ \ \ \textbf{end}\\
      \STATE \ \ \ Find the shortest path \pdenmand \ on \g.
      \STATE \ \ \ Install rule \frl \ to $n$, $\forall \ n \in P_{f_{s,d}}$
      \STATE \ \ \ $TL^{t,b}_{L(u,v)}:=TL^{t,b}_{L(u,v)} \cup f_{s,d}, \forall$ $L(u,v) \in P_{f_{s,d}}$
      \STATE \ \ \ \textbf{Return \pdenmand} 
      \STATE \textbf{Else if} \demand \ leaves the network \textbf{then}
      \STATE \ \ \ $TL^{t,b}_{L(u,v)}:=TL^{t,b}_{L(u,v)} \setminus f_{s,d}, \forall \ L(u,v) \in E$\\
      \textbf{end}
    \end{algorithmic}
\end{algorithm}

\subsection{Failure Recovery Module}
The ``failure recovery module'' takes care of adjusting the link weights based on (2) to find backup path and executing fast rerouting 
to time efficiently recover the data transmission of the affected flows, when link failure occurs.
Additionally, the ``failure recovery module'' carefully sets the priority of flow entries and the timing installing or removing forwarding rules to improve the recovery efficiency. 
Specifically, two sets are maintained in this module.


\begin{itemize}
\item [1.] \ttl \ $:$
We denote \ttl\ as the set of traffic flows on the \link\ in the network. 
The set is updated as long as a traffic flow joins or leaves the network. Also, it is updated when the path of a traffic flow changes.
\item [2.]\tal \ $:$
We denote \tal\ as the set of traffic flows 
on the failure link \lsc \ in the network. 
\tal\ records the traffic flows that need to be rerouted due to the occurrence of the link failure.
Specifically, when the controller receives a packet with a \textbf{LINK REMOVED} message, it records the traffic demand affected by the failure link \lsc. The set is also updated when the controller receives a packet with a \textbf{LINK ADD} message. 
\end{itemize}

Alg. 4 describes the operation of the`` failure recovery module''. 
When the controller receives a \textbf{LINK REMOVED} message, 
it gets the information of the traffic flows affected by \lsc \ in the \ttl. 
Then, the weight of \lsc \ is set to be infinity. 
Before calculating a backup path for each traffic flow affected by a failed link, 
the weights 
\wnew \ for links other than the failed link based on (2) so that
a lightly loaded backup path can be picked for rerouting. 
Further, the SDN controller installs the high priority flow entries \frh \ on the nodes of the backup path aiming to replace the low priority flow entries \frl.
Then,  two corresponding sets \ttl\ and \tal\ are updated in further. 
Moreover, 
the SDN controller directly installs the high priority flow entries, denoted as \frh. In this way, the latency of transmitting the \textbf{Delete} flow operation message can be reduced. The delay in installing new flow entries can also be reduced because the SDN controller does not need to transmit the \textbf{Modify} flow operation message and the communication in obtaining flow entries information from the nodes in the network can be avoided.


When the controller receives the packet with \textbf{LINK ADD} message, 
it checks \tal\ to find the traffic flows routed through their backup paths and then update the residual bandwidth of those follows by Alg. 1.
As a result, those flows can be rerouted by executing Alg. 3.
Furthermore, both the \ttl\ and \tal\ are updated, and the \frh\ of the corresponding nodes are detected.
Note that the priority of the new flow entries  \frl \ is lower than the priority of backup flow entries  \frh. 
Therefore, 
\demand \ is firstly matched by the backup flow entries,  \frh,  before the  \frh \  is deleted. 
After the installation of the low priority flow entries, \frl, on the corresponding node in the network,   
the unnecessary backup flow entries,  \frh, are deleted. 
Through this process, the forwarding path can be switched without the delay of transmitting delete flow instruction from the SDN controller to the corresponding nodes in the network. 
More specifically, the process of \textbf{LINK ADD} makes routing decisions with the assistance of Alg.1 and Alg.3 without worrying about the reroute speed and the communication delay between the controller and the switch. 
In this way, the delay in message exchange between the controller and the APs as well as in computing alternative routes can be decreased.
Further, the efficiency of load balancing can also be improved.

\begin{algorithm}[htbp]
  \caption*{\textbf{Alg. 4:} Failure Recovery Module}
  \begin{algorithmic}[1]
      \STATE Receive a OFPT PORT STATUS packet.
      \STATE \textbf{If} the link state is \textbf{LINK REMOVED} \textbf{then} 
      \STATE \ \ \ Use \lsc \ to find all the affected traffic demand\\
      \ \ \ \demand \ in \ttl
      \STATE \ \ \ $W_{L_{sc}(u,v)} := \infty$
      \STATE \ \ \ $W^{'}_{L_{sc}(u,v)} := \infty$
      \STATE \ \ \ \textbf{Forall} the affected traffic demand \demand \ pass through\\
      \ \ \ \lsc \ \textbf{do}
      \STATE \ \ \ \ \ \ \textbf{Forall} $L(u,v) \in E \setminus L_{sc}(u,v)$ \textbf{do}
      \STATE \ \ \ \ \ \ \ \ \ \textbf{Run Alg. 2}\\
      \ \ \ \ \ \ \textbf{end}
      \STATE \ \ \ \ \ \ Use Dijkstra algorithm to find the backup path \\
      \ \ \ \ \ \ \pdenmand \ on \g \ base on \wnew
      \STATE \ \ \ \ \ \ Install \frh \ to \n, $\forall$ \n     \ $\in$ \pdenmand
      \STATE \ \ \ \ \ \ $AL^{t,b}_{L_{sc}(u,v)}:=AL^{t,b}_{L_{sc}(u,v)} \cup f_{s,d}$
      \STATE \ \ \ \ \ \ $TL^{t,b}_{L(u,v)}:=TL^{t,b}_{L(u,v)} \setminus AL^{t,b}_{L_{sc}(u,v)}\cup$ $f_{s,d}$, $\forall$\\
      \ \ \ \ \ \  \link \ $\in$ \pdenmand\\
      \ \ \ \textbf{end}
      
      \STATE \textbf{Else}
      \STATE \ \ \ Use \lsc \ to find all the affected traffic demand\\
      \ \ \ \demand \ in \tal

      \STATE \ \ \ \textbf{Forall} the affected traffic demand \demand \ pass through\\
      \ \ \ \lsc \ \textbf{do}
      \STATE \ \ \ \ \ \ \textbf{Run Alg. 1 (Line 3-8)}
      \STATE \ \ \ \ \ \ \textbf{Run Alg. 3 (Line 1-9)}
      \STATE \ \ \ \ \ \ $AL^{t,b}_{L(u,v)}:=AL^{t,b}_{L(u,v)} \setminus f_{s,d}$
      \STATE \ \ \ \ \ \ $TL^{t,b}_{L(u,v)}:=TL^{t,b}_{L(u,v)} \setminus f_{s,d}$, $\forall$ \link \ $\in E$
      \STATE \ \ \ \ \ \ $TL^{t,b}_{L(u,v)}:=TL^{t,b}_{L(u,v)} \cup f_{s,d}$, $\forall$ \link \ $\in P_{f_{s,d}}$
      \STATE \ \ \ \ \ \ \textbf{Forall} \n, $\forall$ \n \ $\in V$ \textbf{do}
      \STATE \ \ \ \ \ \ \ \ \ Delete \frh \ in node \n \ \\
      \ \ \ \ \ \ \textbf{end}\\
      \ \ \ \textbf{end}\\
      \textbf{end}
  \end{algorithmic}
\end{algorithm}

\subsection{Description of the \method\ Mechanism}

\begin{figure}[htbp]
\renewcommand{\figurename}{Fig.}
\centerline{\includegraphics[width=0.95\columnwidth]{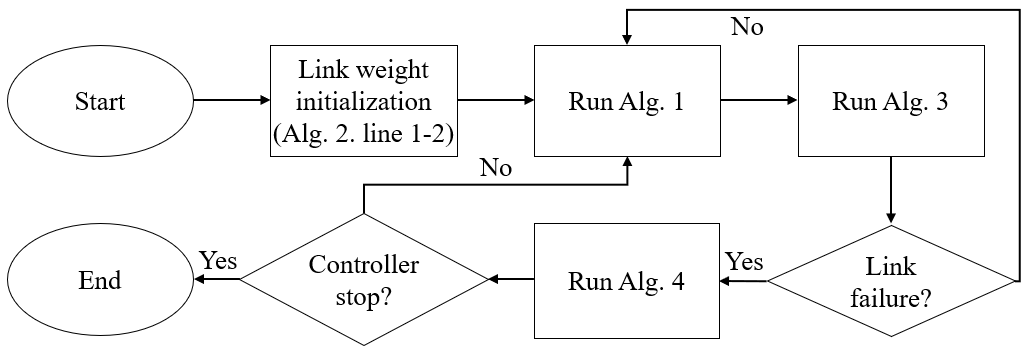}}
\caption{The flowchart of SD-FFR mechanism.} 
\label{flowchat_ALG}
\end{figure}

After properly integrating the four algorithms of the proposed \method\ mechanism, Fig.~\ref{flowchat_ALG} depicts
the flowchart of the \method.
Specifically, when the network starts, 
the \method \ mechanism initializes the link weights by Alg. 2. 
In addition, the Alg. 2 is executed to manage the link weights in the network. 
Following the  Alg. 2, 
information of 
residual link bandwidth is updated by Alg. 1. 
Simultaneously, 
the routing decisions can be made through Alg. 3, when a traffic flow joins the network.  
When the failure occurs, \method \ performs Alg. 4 for failure recovery processing. 

\section{Implementation}
\begin{figure}[htbp]
\renewcommand{\figurename}{Fig.}
\centerline{\includegraphics[width=0.95\columnwidth]{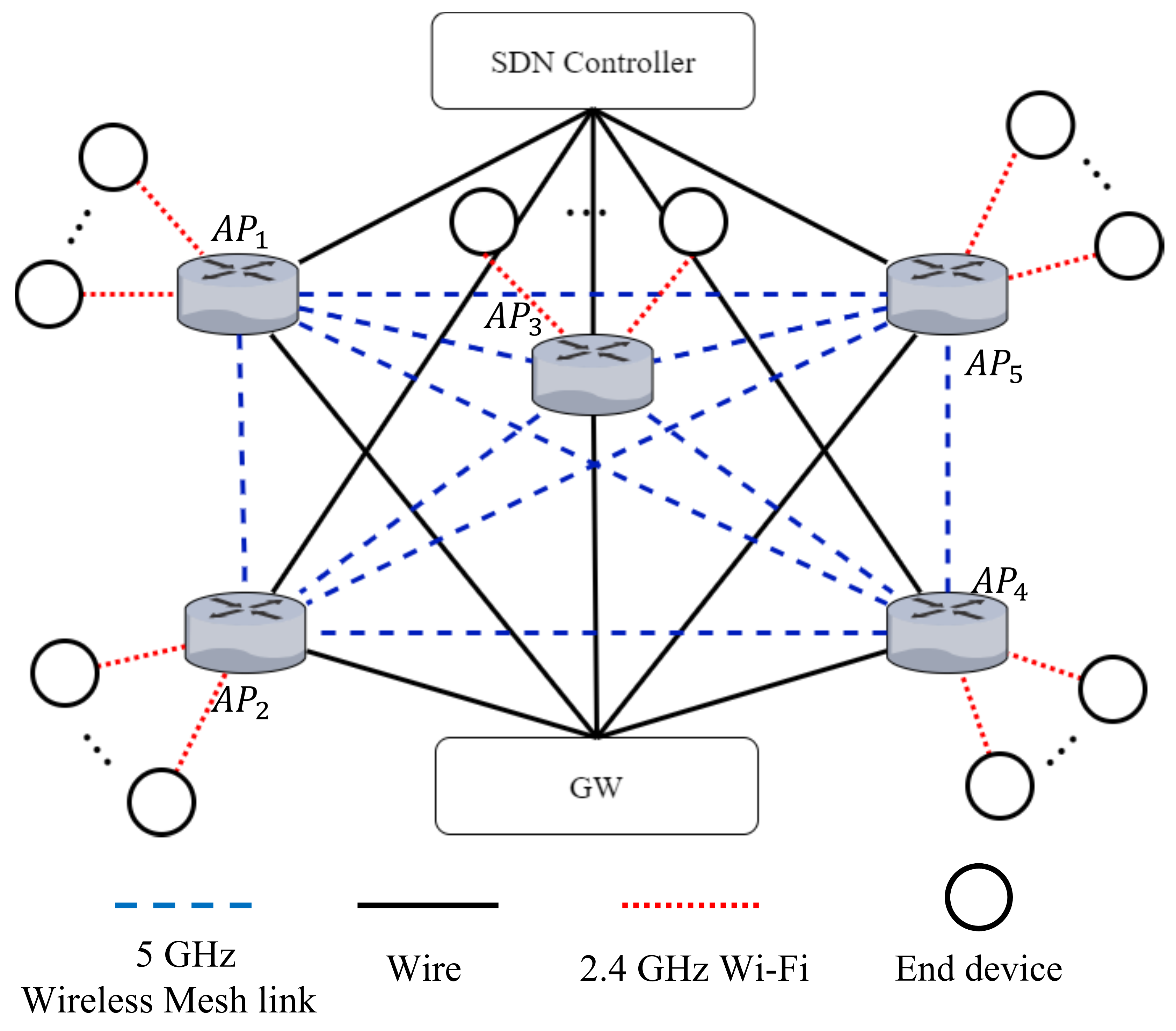}}
\caption{Network environment.} 
\label{fig:NE}
\end{figure}

In this section, we evaluate the behavior of data transmission in SDN assisted dual-band WiFi networks by commercial APs, and apply the measurement for Mininet simulation setup in Section~\ref{sec:eval} aiming to effectively
validate the performance of 
the proposed \method\ in real network environment.
As a result, following Fig.~\ref{fig:NE}, each commercial AP has  Ethernet connectivity to the Internet and the SDN controller,  5 GHz wireless mesh connectivity to neighboring APs and a 2.4 GHz connectivity to
mobile devices.
Further, we enable OpenFlow1.3\cite{pub.1034503230} protocol on the commercial APs (i.e., TP-Link Archer C5 AC175) by
installing both OpenWrt and OpenvSwitch (OVS) on them.
Then, the OVS running on the APs can communicate with the ONOS controller.
We use iPerf and Ping to measure the link bandwidth the data transmission latency under different network conditions.
Specifically, we collect measurements by using wired and wireless link individually which can be further applied for the experimental setup in Mininet.

TABLE~\ref{tab:modeling} presents the measured link bandwidth and data transmission delay in the network formed by the commercial APs.
As shown in TABLE~\ref{tab:modeling}, Ethernet provides the largest link bandwidth (i.e., 100 Mbps) and the smallest data transmission delay (i.e., 0.37ms).
71 Mbps and 47 Mbps are measured through the 5 GHz wireless mesh link and the 2.4 GHz link, respectively. Unsurprisingly, the delay over 5 GHz wireless link
is smaller than the 2.4 GHz wireless link. The results in TABLE~\ref{tab:modeling} will be applied in next section to investigate the network performance in the presence of link failure.

\begin{table}[htbp]
  \centering
  \caption{Mininet modeling.}
  \label{tab-label-1}
\begin{tabular}{|c|c|c|} \hline 
 & Link bandwidth & Delay \\ \hline 
Wired link & 100 Mbps & 0.37 ms \\ \hline
5 GHz wireless mesh link & 71 Mbps & 0.66 ms  \\ \hline
2.4 GHz wireless link & 47 Mbps & 1.8 ms \\ \hline
\end{tabular}
\label{tab:modeling}
\end{table}

\subsection{The Challenge in Link Failure Detection by OVS on Commercial APs}
\label{sec:challenge}
To investigate the behavior of link failure detection and recovery on the commercial APs,
we manually unplug the Ethernet on an AP which in turn cannot access to the Internet.
We found that the OVS cannot instantly receive the link failure information because the synchronization between the OVS and the operating system of the AP is non-realtime. On average, additionally 8 seconds delay is introduced for the ONOS to receive the failure message.
However, this situation does not happen in the Mininet.

\begin{figure}[htbp]
\centerline{\includegraphics[width=0.95\columnwidth]{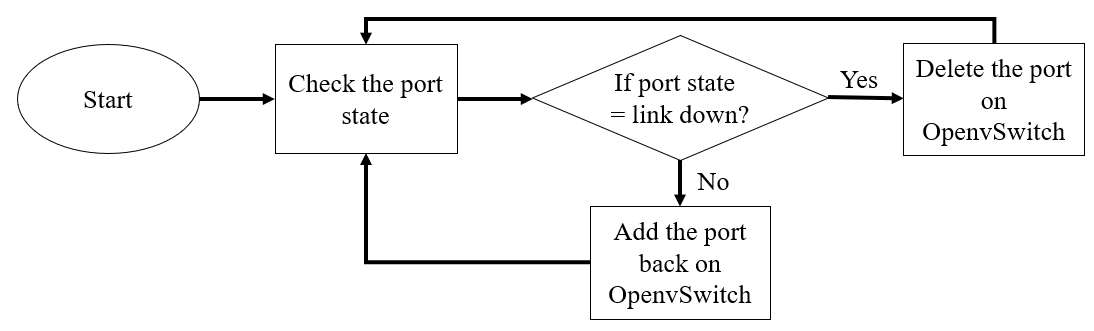}}
\caption{The flowchart of network state detection acceleration module.}  
\label{fig:flowchart}
\end{figure}

\begin{figure}[htbp]
\centerline{\includegraphics[width=0.95\columnwidth]{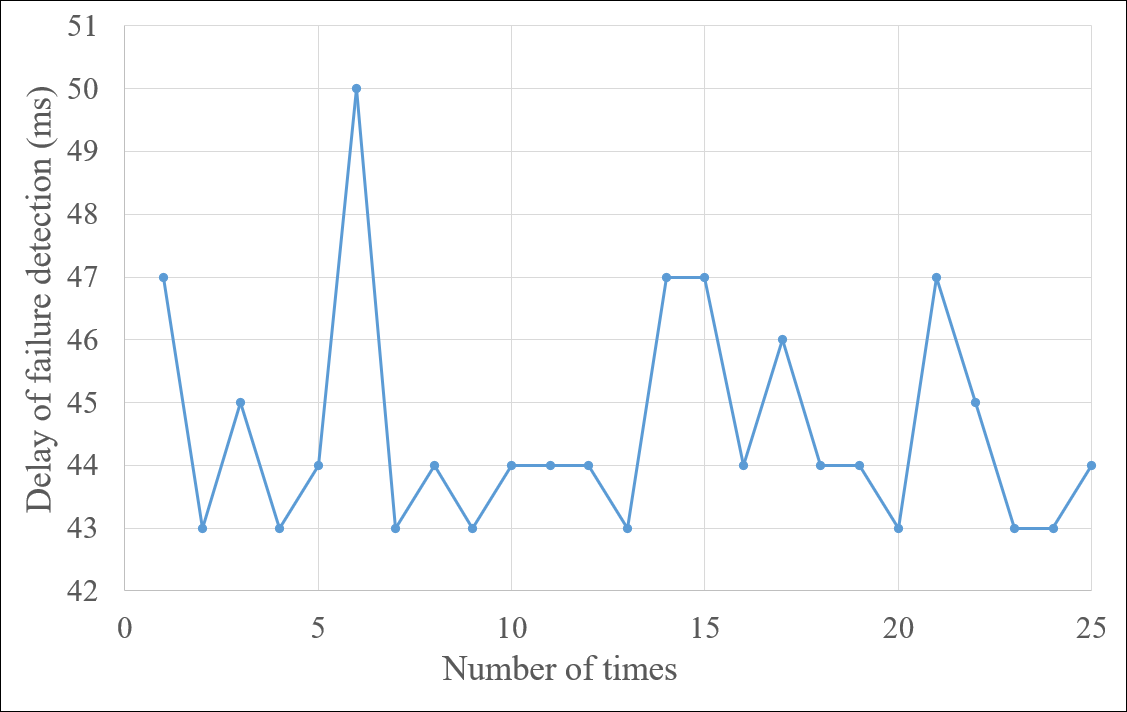}}
\caption{Delay of failure detection.} 
\label{fig:DFD}
\end{figure}

Furthermore,  we design the ``network state detection acceleration module'' shown in Fig.~\ref{fig:flowchart} executing on the data plane, to mitigate the adverse impact on network performance.
Specifically, the AP periodically detects its ports status.
When the AP detects the port status changes, 
the OVS receives Delete or Add command.
Fig.~\ref{fig:DFD} shows 25 measurements of the delay for the AP in detecting whether the port state changes. 
As shown in Fig.~\ref{fig:DFD}, the average delay reduces to 44.6 ms by implementing the 
the ``network state detection acceleration module'' on the AP. 
Then, considering the minimized response time of link failure, we input the 44.6ms delay into the Mininet to evaluate the failure recovery performance of the proposed \method\ in the real network environment.
Additionally, with the assistance of the ``network state detection acceleration module'', the proposed \method\ can support failure recovery more time efficiently in real network environment. 

\section{Evaluation Results}
\label{sec:eval}
\subsection{Experimental Setup}
To evaluate the performance of the proposed \method, we use Mininet to create a network topology present in Fig.~\ref{fig:NE}.
Specifically, the Mininet runs on the computer with Intel i7-8705G CPU and 16 GB RAM.
In the network, five APs not only directly connect to the GW through wired connection but also 
connect with neighboring APs via 5 GHz wireless link. 
Each AP serves provides end devices  wireless connectivity operated at the 2.4 GHz frequency band.
Specifically, we adopt the parameters measured in the commercial WiFi APs shown in TABLE~\ref{tab:modeling}
and the 44.6 ms link failure detection latency described in Section~\ref{sec:challenge}
to configure the network.

Several experiments are conducted to validate the effectiveness of the \method. The goal of the evaluation are twofolds.
First of all, we evaluate the impact of the parameters
\q\ and \alp \ of (1) on the network performance in terms of load balance. 
Furthermore, we investigate the impact network conditions 
on packet loss rate and throughput. 
The method in \cite{8690787} is applied for the comparison, because it considers similar network setup. 

\subsection{The Impact of Parameters on Load Balance}

In this experiments, different combinations of  \q\ and \alp \ are set to investigate the relationships between the link weight function (1) and the achieved
load balance in the normal network state.
Sixteen hosts are introduced on $AP_{1}$ and $AP_{2}$ in the network. 
Each host on $AP_{1}$ performs an iPerf client to communicate with the corresponding iPerf server connecting with $AP_2$, and then
sixteen UDP traffic flows constantly transmit packets with 20 Mbps data rate to the network.
With appropriately picked \q\ and \alp, 
 each WMN link should share three traffic flows while the Ethernet connection handles four traffic flows to avoid link congestion,
 because each AP equips with one Ethernet and four WMN links, 
TABLE~\ref{SPTD} illustrates the number of traffic flows which can be successfully delivered under 
different pairs of \q\ and \alp \ for the performance evaluation. 
As present in TABLE~\ref{SPTD}, when \alp\  $=10$ and \q\ $\in \{0.1, 0.3, 0.5, 0.7\}$ 
\method\ can handle 16 traffic demands. 
Therefore, in the rest of the experiments,  
we set \alp \ $=10$ and  $Q_{0}=0.1$ in calculating the link weights under the normal network state.

\begin{table}[htbp]
  \centering
  \caption{The adjustable parameters $Q$ and $\alpha$ correspond to the number of successfully process traffic demand.}
  \label{SPTD}
\begin{tabular}{|@{}c|c|c|c|c|c|} \hline 
 \diagbox[width=5em,trim=l]{$Q_{0}$}{$\alpha$} & 10 & 20 & 30 & 40 & 50 \\ \hline 
0.1 & 16 & 13 & 10 & 9 & 9 \\ \hline 
0.3 & 16 & 13 & 10 & 9 & 7 \\ \hline 
0.5 & 16 & 13 & 12 & 10 & 7 \\ \hline 
0.7 & 16 & 13 & 12 & 9 & 7 \\ \hline 
0.9 & 15 & 13 & 12 & 10 & 7 \\ \hline 
\end{tabular}
\end{table}

We set $Q_{0}$ to 0.1 and \alp \ to 10 to load balance efficiency in the post-recovery network,
according to TABLE~\ref{SPTD}, $0<2Q_{0}<1$ and $Q_{0}+Q_{1}=1$.
Moreover, instead of introducing 16 iPerf server-client pairs in the previous experiment, only twelves UDP traffic demands whose individual data rate is 20 Mbps.
We disconnect the direct Ethernet connectivity between$AP_1$ and $GW$ as link failure in order to study the performance of the post-recovery network.
TABLE~\ref{LB} illustrates the routing path of each traffic demand before and after fault recovery.
As found in TABLE~\ref{LB}, the traffic demand 1, 3 and 7 are rerouted through $AP_3$, $AP_4$ and $AP_5$, respectively because the link between $AP_1$ and GW fails.
These three flows are evenly redirected through three different paths to avoid the link congestion after failure recovery.

\begin{table}[htbp]
  \centering
  \caption{Load-balance in post-recovery network.}
  \label{LB}
\begin{tabular}{|c|c|c|} \hline 
\textbf{Traffic demand} & \textbf{Original path} & \textbf{Path of post recovery network} \\ \hline 
\textbf{}\textbf{1} & $AP_1-GW-AP_2$ & $AP_1-AP_3-AP_2$ \\ \hline 
\textbf{2} & $AP_1-AP_2$ & $AP_1-AP_2$ \\ \hline 
\textbf{3} & $AP_1-GW-AP_2$ & $AP_1-AP_4-AP_2$ \\ \hline 
\textbf{4} & $AP_1-AP_3-AP_2$ & $AP_1-AP_3-AP_2$ \\ \hline 
\textbf{5} & $AP_1-AP_4-AP_2$ & $AP_1-AP_4-AP_2$ \\ \hline 
\textbf{6} & $AP_1-AP_5-AP_2$ & $AP_1-AP_5-AP_2$ \\ \hline 
\textbf{7} & $AP_1-GW-AP_2$ & $AP_1-AP_5-AP_2$ \\ \hline 
\textbf{8} & $AP_1-AP_2$ & $AP_1-AP_2$ \\ \hline 
\textbf{9} & $AP_1-AP_2$ & $AP_1-AP_2$ \\ \hline 
\textbf{10} & $AP_1-AP_3-AP_2$ & $AP_1-AP_3-AP_2$ \\ \hline 
\textbf{11} & $AP_1-AP_4-AP_2$ & $AP_1-AP_4-AP_2$ \\ \hline 
\textbf{12} & $AP_1-AP_5-AP_2$ & $AP_1-AP_5-AP_2$ \\ \hline 
\end{tabular}
\end{table}

\subsection{The Impact of Network Conditions on Performance}
\subsubsection{The impact of flow rates on packet loss}
\begin{table}[htbp]
  \centering
  \caption{Mean of packet loss rate (\%).}
  \label{PLRM}
\begin{tabular}{|p{0.5in}|p{0.5in}|p{0.5in}|p{0.5in}|p{0.5in}|} \hline 
 & 10Mbps & 15Mbps & 20Mbps & 25Mbps \\ \hline 
Case1 & 0.39 & 0.39 & 0.81 & 1.28 \\ \hline
Case2 & 0.39 & 0.42 & 0.60 & 1.35 \\ \hline
\end{tabular}
\end{table}

\begin{figure}[htbp]
\centerline{\includegraphics[width=0.95\columnwidth]{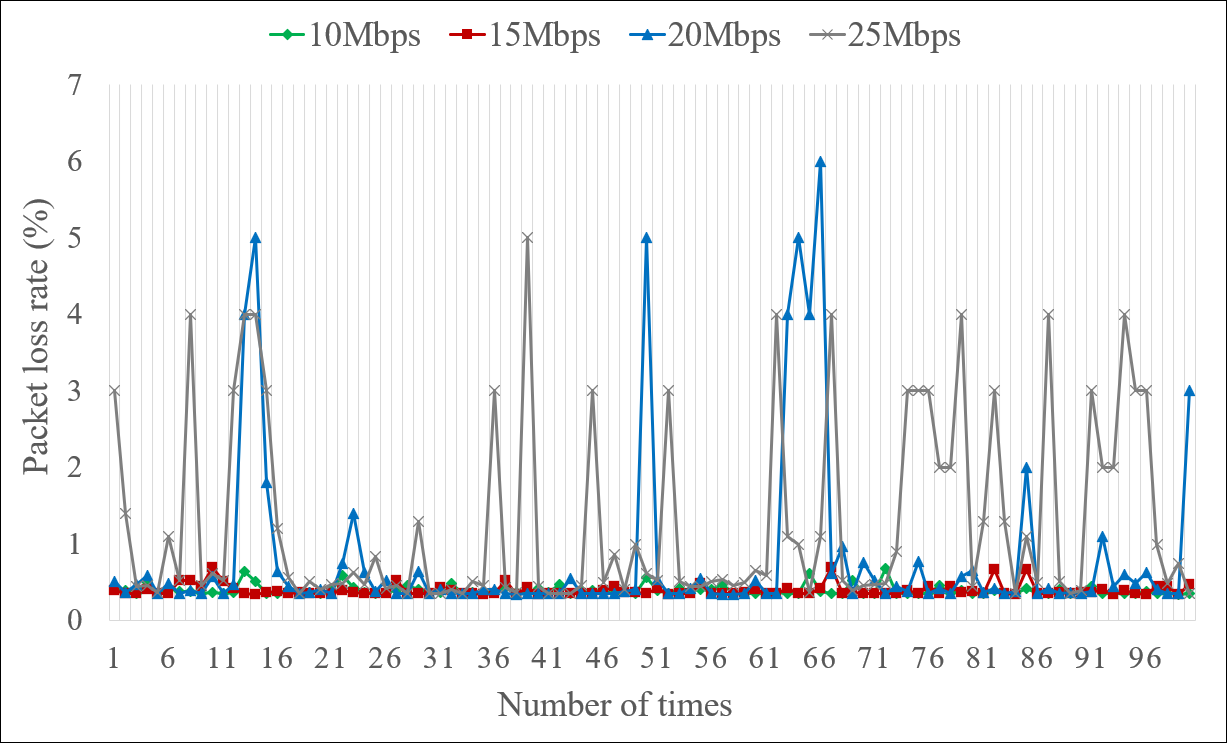}}
\caption{packet loss rate in case 1.}
\label{fig:PLR1}
\end{figure}

\begin{figure}[htbp]
\centerline{\includegraphics[width=0.95\columnwidth]{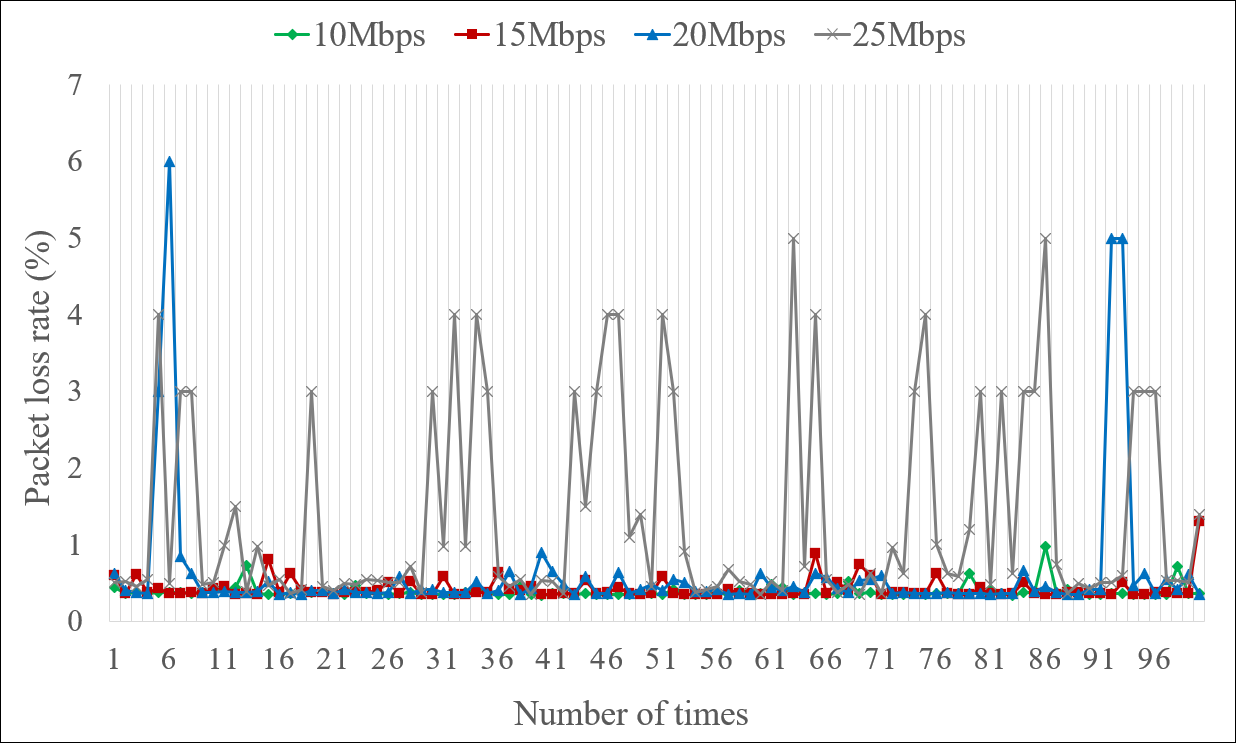}}
\caption{Packet loss rate in case 2.} 
\label{fig:PLR2}
\end{figure}

In this experiment, we consider a traffic flow sent through $AP_1$ and $AP_2$ to the destination.
We used iPerf to generate UDP traffic of 10Mbps, 15Mbps, 20Mbps, and 25Mbps for 30 seconds, respectively and we repeat the measurement 100 times per flow rate  aiming to measure the packet loss rate in the presence of link failure. 
Further, two cases of failure, wired and wireless link failure, are considered. We denote the link failure occurs in the wired connection as case 1 while case 2 denotes the wireless link failure. 
Link failure occurs at the 15th second and only occurs once in each trial.
Fig.~\ref{fig:PLR1}\ -~\ref{fig:PLR2} present the packet loss rate during failure recovery in case 1 and 2 respectively.
The x-axis shows the number of repeated trials and the y-axis indicates the corresponding average packet loss rate.
The packet loss rarely occurs when flow rates are smaller than 20 Mbps in both cases. With the increasing flow rates, the packet loss rate increases, and the traffic flow with 25 Mpbs data rate
approximately experience at most 5\% packet loss. However, on average, the packet loss rate for 25 Mbps data rate is about 1.3\% only as shown in TABLE~\ref{PLRM}.
Since 5 GHz wireless link provides 71 Mbps link bandwidth, the average packet loss rate in both cases are similar as concluded in 
TABLE~\ref{PLRM}. 
Specifically, \method\ time efficiently recovers link failure by rerouting affected traffic flows and carefully plans the operation of failover to reduces the recovery delay. Therefore, the packet loss rate is low.

\subsubsection{The Impact of Link Failure Recovery on TCP Throughput }
\begin{figure}[htbp]
\centerline{\includegraphics[width=0.95\columnwidth]{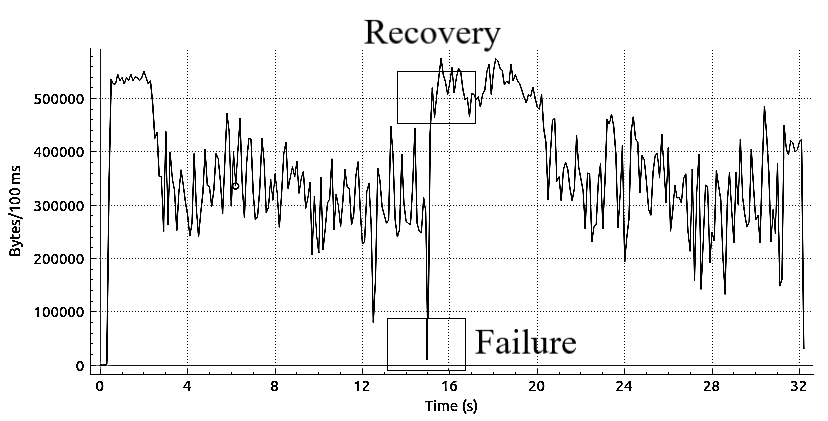}}
\caption{TCP Throughput in case 1.} 
\label{C1T}
\end{figure}

\begin{figure}[htbp]
\centerline{\includegraphics[width=0.95\columnwidth]{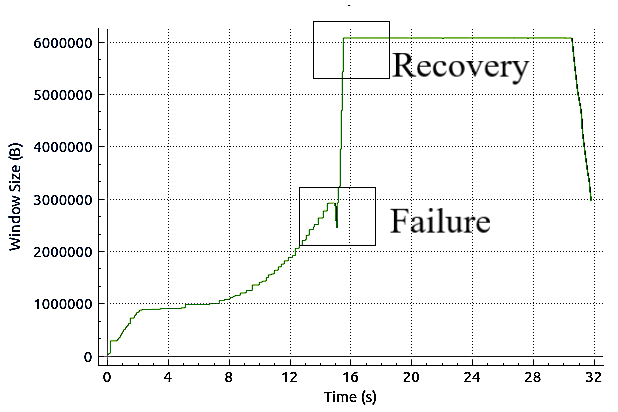}}
\caption{Congestion window size in case 2.} 
\label{C2W}
\end{figure}

In this experiment, we evaluate the impact of link failure recovery on TCP traffic flow following the previous setup.
Hence, the link also fails at the 15th second in both cases.
The obtained TCP throughput and congestion windows in case 1 are shown in Fig.~\ref{C1T}\ -~\ref{C2W}. 
Similarly, Fig.~\ref{C1T}\ -~\ref{C2W} depicts the TCP throughput and the congestion windows in case 2.

It can be found that both the congestion window size and the TCP throughput drops significantly at the 15th second when a link fails in the network. However, the proposed \method\ time efficiently detects the link failure and reroutes the traffic by install the high priority of flow entries during failover. Consequently, the end-to-end performance can be fast recovered.

\begin{figure}[htbp]
\centerline{\includegraphics[width=0.95\columnwidth]{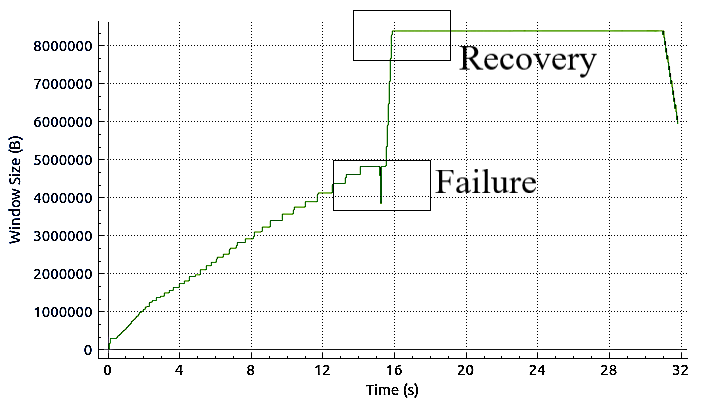}}
\caption{Congestion window size in case 1.} 
\label{C1W}
\end{figure}

\begin{figure}[htbp]
\centerline{\includegraphics[width=0.95\columnwidth]{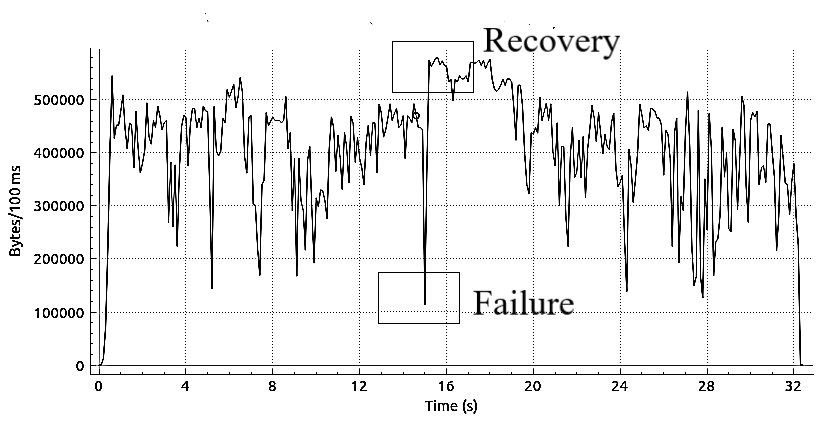}}
\caption{TCP throughput in case 2.} 
\label{C2T}
\end{figure}

\subsection{The impact of failure recovery schemes on network performance}
\begin{figure}[htbp]
\centerline{\includegraphics[width=0.95\columnwidth]{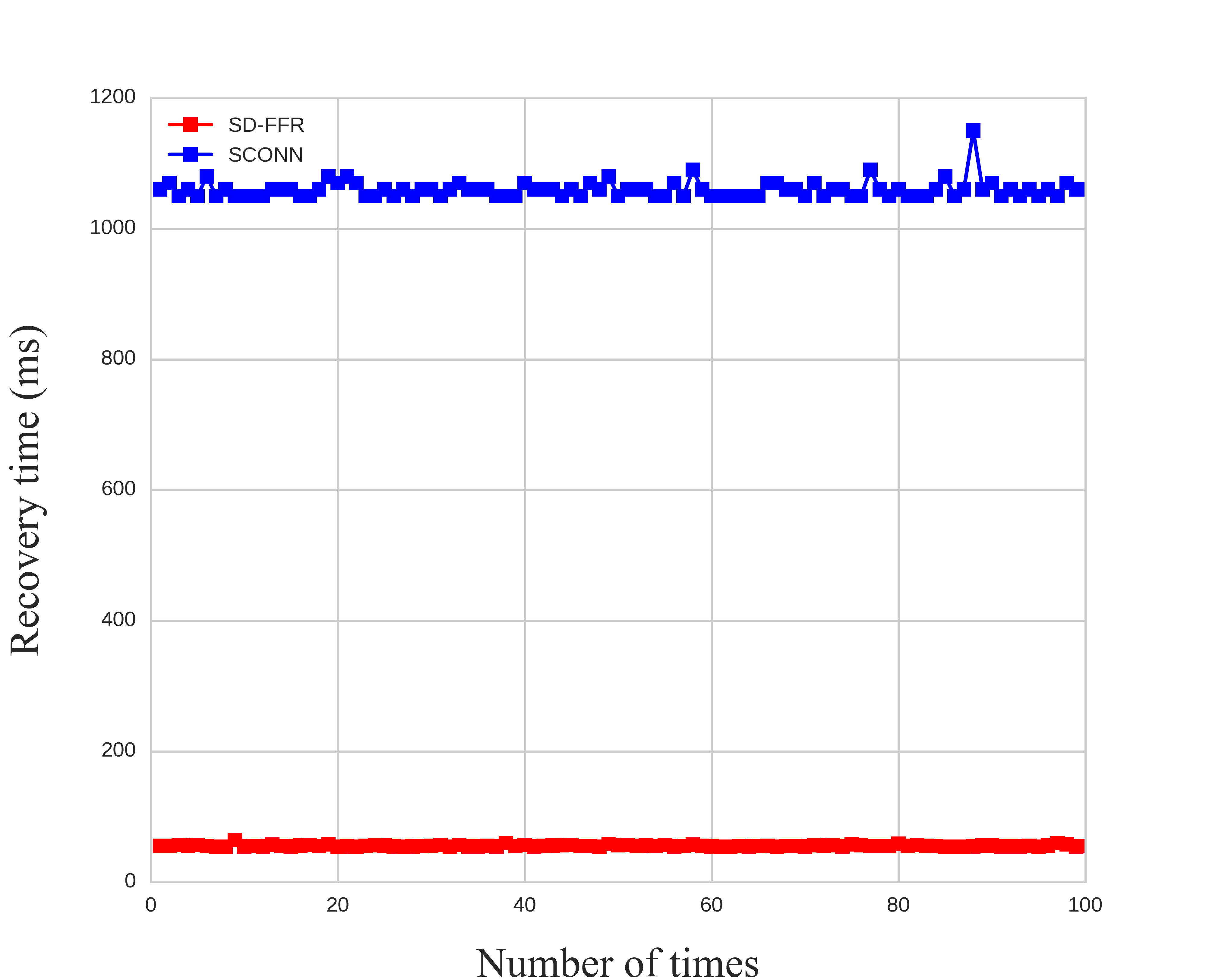}}
\caption{Failure recovery time in case 1.} 
\label{C1R}
\end{figure}

\begin{figure}[htbp]
\centerline{\includegraphics[width=0.95\columnwidth]{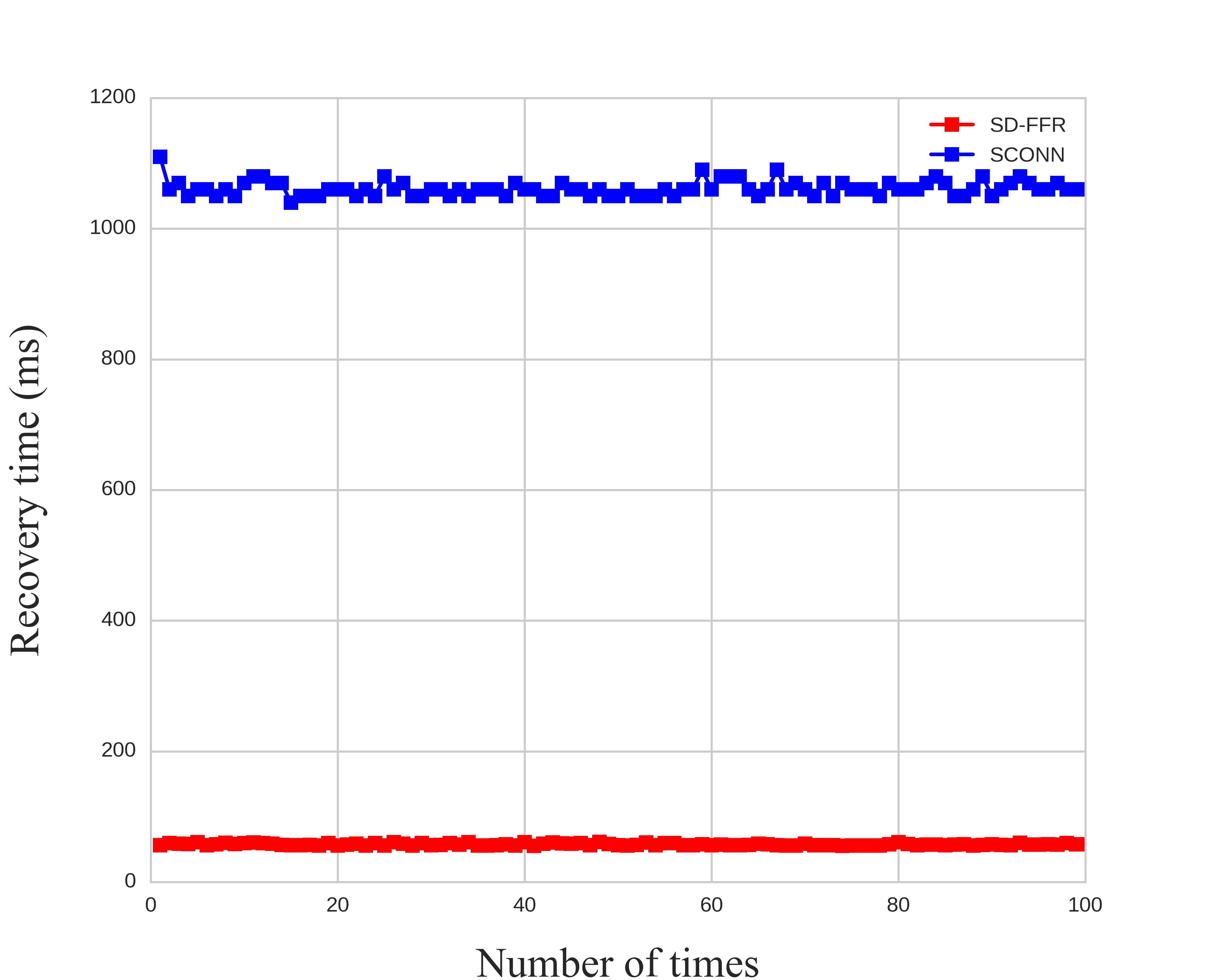}}
\caption{Failure recovery time in case 2.} 
\label{C2R}
\end{figure}

In this experiment, we evaluate the performance of failure recovery through our proposed \method\ and SCONN\cite{8690787} by using Ping.
We introduce the link failure at the 15th second within the 30 seconds experiment, and then repeat the experiment 100 times.
Fig.~\ref{C1R}\ and Fig.~\ref{C2R} plot the link failure recovery time for each trail 
in case 1 and 2, respectively.
The mean and variance of the recovery time are summarized in TABLE~\ref{CBDAOFRT}. 
The proposed \method\ recovers the link failure  18 times faster comparing to SCONN in both cases, since \method\ takes less than 60 ms to complete failure recovery while SCONN
takes at 1000 ms for failure recovery.
It is because \method \ carefully plans the priority of flow entries during failover and performs the installation and deletion of forwarding rules at the right time.
The time in collecting link usage and calculating rerouting paths can be greatly saved which in turn reduces the recovery delay significantly.
SCONN does not have these considerations, and thus the recovery time is high. 

\begin{table}[htbp]
  \centering
  \caption{Compare between different approaches on failure recovery time.}
  \label{CBDAOFRT}
\begin{tabular}{|c|c|c|} \hline 
Average recovery time and STD & Case 1 & Case 2 \\ \hline 
\method \ & \makecell[c]{Mean: 55.5ms\\STD: 1.48} & \makecell[c]{Mean: 57.6ms\\STD: 1.59} \\ \hline
SCONN\cite{8690787} & \makecell[c]{Mean: 1060ms\\STD: 0.01}& \makecell[c]{Mean: 1061ms\\STD: 0.01}\\ \hline
\end{tabular}
\end{table}

\section{Conclusion}
In this paper, we propose a \method \ mechanism to improve the network reliability in the automatic warehouse.
With \method, the SDN controller reroutes the traffic to the backup wireless mesh network to react to the occurrence of link failure. 
Specifically, 
\method \ carefully sets the priority of the flow entries on each nodes in the network to minimize  
the appropriate operation time installing and deleting the corresponding forwarding rules of the traffic flows influenced by the link failure. 
Therefore, the link failure can be time efficiently recovered. 
Considering that the traffic load in the automatic warehouse may be unevenly balanced, 
the \method \ mechanism also design strategy in balancing traffic loads no matter link failure happens or not. 

We evaluate the proposed \method \ mechanism on Mininet, and the results show the effectiveness of \method\ in improving the network performance. Specifically, in the Mininet evaluations, the parameters of link bandwidth and latency in recovering link failure are configured, according to the measurement by commercial APs to study the impact of applying \method\ in the real network.
In the future, we will 
design a hybrid failure recovery scheme integrating the active and passive failure recovery strategies in order to improve our proposed mechanism further.

\small
\bibliographystyle{IEEEtran}
\bibliography{SD-FFR_j_final}

\end{document}